\documentclass[conference]{IEEEtran}
\IEEEoverridecommandlockouts
\usepackage{cite}
\usepackage{amsmath,amssymb,amsfonts}
\usepackage{algorithmic}
\usepackage{graphicx}
\usepackage{multirow}
\usepackage{textcomp}
\usepackage{xcolor}
\def\BibTeX{{\rm B\kern-.05em{\sc i\kern-.025em b}\kern-.08em
    T\kern-.1667em\lower.7ex\hbox{E}\kern-.125emX}}
\begin{document}

\title{Improving Development Practices through Experimentation: an Industrial TDD Case}

\author{\IEEEauthorblockN{Adrian Santos, Jaroslav Spisak*, Markku Oivo and Natalia Juristo**}
\IEEEauthorblockA{M3S ITEE, University of Oulu, Finland\\
*Play among friends (Paf), Finland \\
**Universidad Polit\'ecnica de Madrid, Spain \\
\{adrian.santos.parrilla/markku.oivo\}@oulu.fi, jaroslav.spisak@paf.com, natalia@fi.upm.es}}

\maketitle

\begin{abstract}

Test-Driven Development (TDD), an agile development approach that enforces the construction of software systems by means of successive micro-iterative testing coding cycles, has been widely claimed to increase external software quality. In view of this, some managers at Paf---a Nordic gaming entertainment company---were interested in knowing how would TDD perform at their premises. Eventually, if TDD outperformed their traditional way of coding (i.e., YW, short for Your Way), it would be possible to switch to TDD considering the empirical evidence achieved at the company level. We conduct an experiment at Paf to evaluate the performance of TDD, YW and the reverse approach of TDD (i.e., ITL, short for Iterative-Test Last) on external quality. TDD outperforms YW and ITL at Paf. Despite the encouraging results, we cannot recommend Paf to immediately adopt TDD as the difference in performance between YW and TDD is small. However, as TDD looks promising at Paf, we suggest to move some developers to TDD and to run a future experiment to compare the performance of TDD and YW. TDD slightly outperforms ITL in controlled experiments for TDD novices. However, more industrial experiments are still needed to evaluate the performance of TDD in real-life contexts. 

\end{abstract}


\section{Introduction}
TDD is an agile development approach that enforces the construction of software systems by means of successive micro-iterative testing-coding cycles \cite{beck2003test}. These micro-iterative testing-coding cycles are, according to its proponents \cite{beck2003test}, the main reason behind TDD's superiority over traditional approaches (e.g., Waterfall) on external quality\footnote{External quality is usually considered in the literature on TDD as the number of tests that pass from a battery of tests specifically built for testing the application under development \cite{kollanus2010test, shull2010we, causevic2011factors, rafique2013effects}. For simplicity's sake, along this article we will refer to quality and external quality interchangeably.}: while testing \textit{before} coding forces developers to think ahead about the structure of the code---contrary to traditional approaches where the structure "emerges" after coding---the \textit{iterative} nature of TDD forces developers to re-factor (i.e., polish the code to make it more consistent and maintainable over time) and increase the functionality of the system in \textit{small} iterations---contrary to traditional approaches where software systems are generally built up-front and then tested. In turn, this should increase the consistency of the whole system and make each of its functionalities less error-prone over time.

Not just TDD's proponents claim its superiority over traditional approaches on external quality: a large body of empirical research (mostly case studies and surveys) back-up such claims also \cite{kollanus2010test, shull2010we, causevic2011factors, rafique2013effects, makinen2014effects, munir2014considering, bissi2016effects}. For example, according to the case studies conducted so far, TDD outperforms control approaches in considerable ways: from increases in external quality as low as 18\% \cite{mcdaid2008test}, to as high as 50\% \cite{damm2006results}. However, a different picture is provided by controlled experiments: negative, neutral or positive results emerge depending upon the control approach being compared (i.e., ITL or the Waterfall) or the environment where TDD is evaluated (i.e., industry vs. academia) \cite{rafique2013effects, kollanus2010test}. Besides, despite the alleged benefits of \textit{industrial experiments} (e.g., providing cause-effect claims on technology performance in realistic settings \cite{sjoberg2002conducting}, generating and validating theories in industry-relevant contexts \cite{falessi2017empirical}, etc.), almost none of the experiments have been run in industry so far. This led Munir et al. to claim \cite{munir2014considering}: \textit{"...strong indications are obtained that external quality is positively influenced, which has to be further substantiated by industry experiments..."}.

Encouraged by the results achieved by TDD in the literature, Paf's\footnote{https://www.paf.com/} managers were interested in evaluating the extent to which TDD would perform at their premises. Eventually, if TDD outperformed their current development practices, it may be possible to move their software development team to TDD.

Along this paper we aim to answer two \textbf{research questions}:
\begin{itemize}
    \item{\textbf{RQ1:} Does TDD outperform current development practices at Paf with regard to external software quality?}
    \item{\textbf{RQ2:} Do Paf's results agree with those of previous experiments on TDD?}
\end{itemize}

To answer these research questions, we conducted a three-day seminar on TDD at Paf with an embedded experiment. In the experiment we evaluated the extent to which the amalgamation of the development practices followed by Paf's developers (i.e., YW), ITL and TDD performed on external quality. Then, we went over the secondary studies conducted so far on TDD to identify the controlled experiments that had been run. We made several \textbf{findings}: 
\begin{itemize}
    \item{TDD outperformed YW and ITL to a small extent at Paf. }
    \item{Paf's results cannot be compared with those of any other industrial experiment as the sole experiment conducted so far on TDD in industry was not able to finally assess external quality.}
    \item{TDD slightly outperforms ITL for TDD novices in controlled experiments.}
\end{itemize}

The main \textbf{contributions} of this paper are an \textit{evaluation of the performance of TDD against YW and ITL in an industrial experiment} with regard to external quality and \textit{a comparison and meta-analysis} of Paf's results with those of already published \textit{controlled experiments}.\footnote{In this study we use Wohlin et al. \cite{wohlin2012experimentation} and Juristo et al. \cite{juristo2013basics} definitions of controlled experiment to consider a primary study as a valid controlled experiment.}

Along this study we argue that while case studies and surveys show that TDD clearly outperforms control approaches, such claims seem not supported by controlled experiments. In addition, despite the long years of research on TDD, little is yet known on how TDD performs in industrial experiments \cite{offutt2018don}. In view of this, we conclude:
\\
\\
\noindent\fbox{
  \parbox{8cm}{
    \textbf{Take-away messages}
    \begin{itemize}
        \item{We cannot suggest Paf to immediately adopt TDD in view of the results achieved: despite TDD outperformed their current development practices, the difference in performance was minimum.}
        \item{Due to the encouraging results achieved with TDD at Paf, we recommend Paf's managers to move some of their developers to TDD and eventually, after developers are acquainted with sufficient experience with it, to conduct a new experiment to make a decision on whether to adopt TDD in practice.}
        \item{Differences between experiments and case studies or surveys' results with regard to external quality may be due to the lack of familiarity of experiments' participants with the TDD process. However, this finding should be further substantiated, as yet there is a scarce number of industrial experiments evaluating the performance of TDD on external quality.}
    \end{itemize}
   }
}
\\
\\
\textbf{Paper organization}. In Section \ref{related} we provide the related work of this study. In Section \ref{motivation} we motivate the experiment that we run at Paf. Then, in Section \ref{experiment} we outline the characteristics the experiment, while in Section \ref{analysis} we undertake its analysis. Afterwards, we meta-analyze together Paf's results and those achieved in other controlled experiments in Section \ref{comparing}.  We discuss our findings in Section \ref{discussion}. We outline the threats to validity of our study in Section \ref{threats}. Finally, we show the conclusions of this study in Section \ref{conclusion}.

\section{Related Work}
\label{related}

\subsection{Secondary Studies on TDD}

In this section we go over the secondary studies conducted so far on TDD \cite{kollanus2010test, shull2010we, causevic2011factors, rafique2013effects, makinen2014effects, munir2014considering, bissi2016effects} to provide an overview of the primary studies that evaluate the performance of TDD with regard to external quality. Table \ref{summary_tdd} shows the number of primary studies identified in each secondary study, the classifications provided by secondary studies' authors to categorize them and their results.

\begin{table*}[t!]
\small
\begin{center}
\caption{Secondary studies' results on the performance of TDD on external quality.}
\label{summary_tdd}
\begin{tabular}{ l | l | l| l| l} \hline \hline
\textbf{Secondary Study} & \textbf{Classification provided} & \textbf{Positive} & \textbf{Inconclusive} & \textbf{Negative} \\ \hline
\multirow{3}{*}{Kollanus \cite{kollanus2010test}} & Experiments & 2 & 4 & 1 \\
 & Case studies & 9 & 1 & - \\
& Others & 5 & - & - \\ \hline
\multirow{3}{*}{Shull et al. \cite{shull2010we}} &Controlled experiment & 1 & 3 & 2 \\
& Pilot study & 6 & 2 & - \\
&Industrial Use & 6 & 1 & - \\ \hline
\multirow{2}{*}{Causevic et al. \cite{causevic2011factors}} &Experiments & 2 & 2 & 1 \\
& Case Studies & 11 & - & - \\ \hline
\multirow{2}{*}{Rafique et al. \cite{rafique2013effects}} &Standardized meta-analysis & 4 & - & 7 \\
& Unstandardized & 17 &  & 7 \\ \hline
Makinen et al. \cite{makinen2014effects} & Academia/Industry & 5 & 2 & 1 \\ \hline
\multirow{4}{*}{Munir et al. \cite{munir2014considering}} &High rigor/High relevance & 7 & - & - \\
 & Low rigor/High relevance & 3 & - & - \\ 
 & High rigor/Low relevance & 3 & 6 & 1 \\
 & Low rigor/Low relevance & 1 & 1 & - \\ \hline
\multirow{7}{*}{Bissi et al. \cite{bissi2016effects}} &Experiment/Academia & 4& 1& 1\\
& Experiment/Industry & 3& - & - \\
& Case Study/Academia & 1& - & -\\
& Case Study/Industry & 4& - & - \\
& Questionnaire/Academia & 1& - & -\\
& Questionnaire/Industry & 1&- & - \\
& Simulation/Industry & 1&- & -\\ \hline
\end{tabular}
\end{center}
\end{table*}

Kollanus identified a total of 22 empirical studies evaluating the effects of TDD on external quality \cite{kollanus2010test}. Among the seven experiments he identified, two showed benefits, four showed no difference and just one showed detrimental effects. The rest of studies (ten case studies and five surveys) showed an increase in quality, while just one showed inconclusive results. In view of this evidence, Kollanus ends up claiming that \textit{"most of evidence suggests that TDD improves external quality"}.

Shull et al. identified a total of 21 studies. Afterwards, they classified all studies into three categories: (1) controlled experiments: usually academic experiments; (2) pilot studies: \textit{small studies} that were conducted under realistic conditions; and (3) industrial studies: \textit{large real-life} projects undertaken under real commercial pressures \cite{shull2010we}. Again, conflicting results emerged with regards to quality: while controlled experiments showed either no difference or a decline in quality, most pilot studies and industrial studies showed positive results. Despite the conflicting evidence, Shull et al. claim: \textit{"moderate evidence exists for the argument that TDD improves the code's external quality"}.

Causevic et al. identified a total of 16 studies including experiments, case studies and surveys \cite{causevic2011factors}. 13 studies (including experiments and case studies) claim benefits, two claim inconclusive results and only one shows detrimental effects. In view of this evidence, Causevic et al. suggest that \textit{"code quality improvement... is one of the reasons why TDD is gaining interest"}. 

Rafique et al. identified a total of 17 studies \cite{rafique2013effects}. 11 out of those 17 studies were analyzed by means of \textit{standardized meta-analysis} \cite{borenstein2011introduction}. The meta-analysis showed conflicting evidence: while TDD outperformed the Waterfall, TDD underperformed ITL. All 17 studies were then aggregated with a less formal approach (i.e., an unstandardized analysis according to the authors \cite{rafique2013effects}). With this last analysis approach, small improvements in quality were observed with TDD. 

Makkinen et al. identified a total of 8 studies assessing external quality\footnote{External quality's definition was different in this study than in the rest: external quality was evaluated from qualitative data (e.g., clients interviews) rather than quantitative data (e.g., by means of assertions). Thus, the results of this study on the response variable "defect density" are taken here, as this response variable captures what we---and most authors---claim external quality to be in the TDD literature: the number of tests passed from a battery of tests.}: four industrial case studies show improvements and four academic experiments show conflicting results: one positive, one negative and two neutral \cite{makinen2014effects}.

Munir et al. identified a total of 22 studies assessing quality (including experiments, case studies and surveys) and then categorized them into a 2x2 grid \cite{munir2014considering}. Each cell of the grid represented a certain level of relevance (high or low) and rigor (high or low). While rigor represented the degree to which authors follow best practices for reporting and conducting the study, relevance represented the extent to which the study's results may be relevant to industrial practice. High rigor and high relevance studies---the most relevant for the practitioners according to the authors \cite{munir2014considering}---favour TDD over the control development approaches. None study in such category is a controlled experiment. Conflicting results were obtained in the rest of categories. Munir et al. claim in the abstract: \textit{"strong indications are obtained that external quality is positively influenced, which has to be further substantiated by industry experiments..."}. 

Finally, Bissi et al. \cite{bissi2016effects} identified a total of 17 studies. Studies were divided according to its research method: experiments, case studies, questionnaires and simulations. Overall, seven experiments, five case studies, two questionnaires and one simulation showed that TDD increases external quality. One experiment showed inconclusive results and another negative effects. Bissi et al. end up claiming: "about 88\% of the total reported a significant improvement in external software quality".

Long story short, most empirical studies conducted so far on TDD are either case studies or surveys. In most of them, TDD outperforms control approaches. However, experiments show conflicting results: while TDD outperforms control approaches in some, the opposite happens in others. Are experiments' results consistent at least \textit{within} control approaches or \textit{within} contexts (i.e., academia vs. industry)?

\subsection{Controlled Experiments on TDD}

Again, we recurred to the secondary studies on TDD with the aim of gathering a list of all the controlled experiments evaluating external quality. Along this section we follow Wohlin et al.'s \cite{wohlin2012experimentation} and Juristo et al.'s \cite{juristo2013basics} definitions to consider a primary study as a valid \textit{controlled experiment}. This is, we consider a primary study as a controlled experiment whenever at least \textit{two different treatments} (e.g., TDD vs. Waterfall, or TDD vs. ITL) are assessed on a \textit{common dependent variable} (e.g., external quality) in a \textit{controlled environment} (e.g., industrial, or academic laboratory settings), and subjects are assigned to treatments either \textit{completely at random, or by stratification}. Thus, we leave out of this category primary studies that fail to meet some of this criteria. For example, we do not consider experiments those primary studies that:
\begin{itemize}
    \item{\textbf{Allow subjects to work from home or along multiple days} (e.g., \cite{edwards2004using}). We get rid of these studies as they lack a controlled environment, and thus, external factors may impact results (e.g., subjects working from home may be helped by colleagues or be interrupted by unknown factors).}
    \item{\textbf{Evaluate the effects of TDD and another technology jointly} (e.g., \cite{george2004structured}). We do not consider these studies as the effects of TDD and other technologies are confounded and thus, it is not possible "disentangling" the effects of TDD on results.}
    \item{\textbf{Evaluate the effects of TDD on a single subject} (e.g., \cite{madeyski2007impact}). We do not consider such studies because subjects are not randomly assigned to treatments and thus, results are largely dependent upon the skills, experiences and preferences of a single developer.}
    \item{\textbf{Evaluate a variant of TDD instead of TDD} (e.g., \cite{rahman2007applying}). We get rid of those studies as results obtained with such technology may be different from those obtained with TDD.} 
\end{itemize}

Table \ref{experiments_tdd} shows the controlled experiments that we identified in the secondary studies, divided by control approach (i.e., Waterfall vs. ITL), result (i.e., Positive, Negative and NA for "Not Available") and context (i.e., academia vs. industry).\footnote{Gupta et al.'s experiment \cite{gupta2007experimental} is reported twice as it reports the effects of TDD against the Waterfall in two different tasks separately.} As it can be seen in Table \ref{experiments_tdd}, only one controlled experiment was run in industry (i.e., Geras et al. \cite{geras2004prototype}). Unfortunately, Geras et al. were unable to assess the effects of TDD on external quality, as according to them, all subjects obtained the maximum quality score that could be achieved in both the TDD and the control group \cite{geras2004prototype}. With regard to the rest of experiments, even though most show detrimental effects, in some TDD outperforms control approaches. 

\begin{table}[h!]
\centering
\caption{Controlled experiments on TDD.}
\label{experiments_tdd}
\begin{tabular}{p{2cm}|c|c|c} \hline \hline
& \textbf{Positive} & \textbf{Negative} & \textbf{NA} \\ \hline 
\textbf{Waterfall} & \cite{gupta2007experimental} & \cite{wilkerson2012comparing}\cite{gupta2007experimental}\cite{mueller2002experiment}\cite{zieliriski2005preliminary} & \\ 
\textbf{ITL} & \cite{panvcur2011impact} & \cite{pancur2003towards}& \cite{geras2004prototype}* \\ \hline
\multicolumn{4}{l}{* Industrial experiment.}
\end{tabular}
\end{table}

In view of the scarcity of industrial experiments conducted so far, and the conflicting evidence obtained across research methods (i.e., surveys, case studies and controlled experiments), types of subjects (i.e., professionals and students) and control approaches (i.e., ITL or Waterfall), we could not provide a clear answer to Paf's managers with regard to the expected performance of TDD on their premises. In view of this, we proposed them to conduct their own experiment. This way, they will be able to get specifically tailored answers to their technological environment and developers' characteristics.

\section{Experimentation in Software Industry}
\label{motivation}

According to Software Engineering (SE) researchers \cite{sjoberg2002conducting, sjoberg2003challenges}, it is within the main goals---if not the ultimate criterion for success \cite{kitchenham2004evidence}---of any empirical research the transference and thorough consideration of research results in industrial practice. Thus, they claim, SE researchers should strive to conduct industrial studies with the aim of making results applicable to practitioners \cite{kitchenham2004evidence, dyba2005evidence}. 

Among all study types (i.e., case studies, surveys, etc.), industrial experiments are seen as one of the most desirable approaches to evaluate the performance of new technologies in realistic environments \cite{dyba2005evidence, kitchenham2004evidence}. In Sjoberg et al.'s words \cite{sjoberg2002conducting}: \textit{"software engineering researchers should apply for resources enabling expensive and realistic software engineering experiments"}.

The Experimental Software Engineering Industrial Laboratory (ESEIL Project\footnote{For more information, visit http://www.softwareindustryexperiments.org/}) is to the best of our knowledge, the first SE project conducting experiments on TDD across multiple software industries. Instead of just running experiments, we embed them within training courses. This way, experiments are not seen as a cost by companies, but instead, as an investment in which companies receive training in a technology of interest, and at the same time, a timely evaluation on the performance of the technology in company-relevant scenarios. Even though this approach has its own shortcomings (e.g., as experiments are embedded within training courses, professionals enrolling in training courses are novices in the technology under evaluation), we believe, it facilitates the collaboration between academia and industry towards a first-step evaluation of technologies in real-life contexts.

Along the next section, we outline the characteristics of the experiment on TDD that we run at Paf.

\section{Experiment}
\label{experiment}
\subsection{Variables and Research Questions}

The main independent variable within the experiment is \textbf{development approach}, with \textit{YW}, \textit{ITL} and \textit{TDD} as levels. YW is the "amalgamation" of all the development practices followed by Paf's developers (i.e., each developer applies its usual way of coding). ITL is defined as the reverse-order approach of TDD following Erdogmus et al. \cite{erdogmus2005effectiveness}. 

A second independent variable within the experiment is \textbf{experimental task}\footnote{The specification of all the tasks can be found in the following link: http://www.grise.upm.es/Appendix/090618/Tasks.pdf.}. Experimental task has three levels:
\begin{itemize}
    \item{\textit{Bowling-Score Keeper (BSK)} is a modified version of Robert Martin's Bowling Scorekeeper \cite{martin2001advanced}. The goal of BSK is to calculate the score of a single bowling game. BSK is algorithm-oriented, does not involve the creation of a graphic user interface (GUI), and does not require prior knowledge of bowling scoring rules to be developed (as this knowledge is embedded within the specification). We partitioned the specification of BSK into 13 fine-grained sub-tasks. Each sub-task of BSK contained a short, general description, a requirement specifying what that sub-task is supposed to do, and an example consisting of an input and the expected output. We selected BSK as it is one of the most used tasks in TDD experiments \cite{tosun2017industry}.}
    \item{\textit{Mars-Rovers (MR)} is a programming exercise that requires the development of a public interface for controlling the movement of a fictitious vehicle on a grid with obstacles. In particular, the implementation of MR leverages an NxN matrix data structure, where each matrix cell may either contain or not an obstacle through which the vehicle cannot cross. Obstacles are placed within the grid by parsing some initialization commands. The movement of the vehicle is controlled via parsing the standard console input. MR is a popular exercise used by the agile community to teach and practise unit testing. We selected MR as it similarly complex as BSK \cite{tosun2017industry}. }
    \item{\textit{Spread-Sheet (SS)} is a programming exercise that requires the development of a basic spreadsheet without GUI able to perform basic operations on integers and strings: addition, substraction, multiplication, division, module and concatenation. The spreadsheet is organized in rows and columns (similar to MS Excel\footnote{https://products.office.com/es/excel}) and shall follow Excel conventions towards the creation of basic formulas. We selected SS due to its intuitiveness and the relatively well-known functionality of spreadsheets. }
\end{itemize}

The dependent variable within the experiment is \textbf{external quality}. We measure external quality as the percentage of test cases that successfully pass from a battery of test cases that we built to test participants' solutions. Specifically, we measure external quality as:
$$ QLTY =\frac{\#Test~Cases(Pass)}{\#Test~Cases(All)}*100 $$ 

We built a total of 48, 52 and 43 test cases for testing BSK, MR, and SS, respectively. These test cases were also used in a previous experiment that we run on TDD \cite{tosun2017industry}.

One main \textbf{research question} drives Paf's experiment:
\begin{itemize}
    \item{Do YW, ITL and TDD perform similarly in terms of external quality?}
\end{itemize}

\subsection{Subjects}

\begin{figure}[h!]
\includegraphics[width=9cm,keepaspectratio]{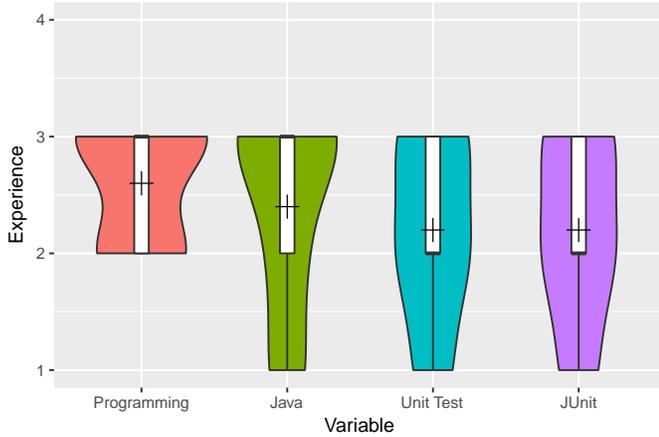}
\caption{Experiences box-plot and violin-plot} \label{fig0}
\end{figure}

A three-day seminar on TDD was conducted at Paf. A total of 15 subjects attended the seminar. Subjects were handed a survey some days before the seminar. The survey contained a series of self assessment questions that followed the same template: \textit{"How would you rate your experience with X"?}, being \textit{X} either TDD, programming, unit testing, Java or JUnit (the programming language and the testing tool used during the experiment, respectively). Each question could be answered in an ordinal-scale (i.e., inexperienced, novice, intermediate and expert). Figure \ref{fig0} shows the box-plot and violin-plot of the participants' experiences.\footnote{Inexperienced, novice, intermediate and experts corresponding to 1, 2, 3 and 4, respectively.} As no subject had any prior experience with TDD, its corresponding box-plot and violin-plot are not shown in Figure \ref{fig0}. 

As it can be seen in Figure \ref{fig0}, Paf's participants classed themselves as intermediate experienced programmers with intermediate experience in Java and a lower experience in unit testing and JUnit. Besides, while most participants had some previous experience with programming, some classed themselves as completely inexperienced with either Java, unit testing or JUnit. As a summary, Paf's participants are an \textit{heterogeneous sample of TDD novices with intermediate experience in programming and Java, and a lower experience in unit testing and JUnit}.

\subsection{Experimental Design and Settings}

An experiment was embedded within the seminar conducted at Paf. Table \ref{schedule} shows the design of the experiment.

\begin{table}[h!] \centering 
  \caption{Experimental design: Paf.} 
  \label{schedule} 
\begin{tabular}{llll} \hline \hline 
\textbf{Group} & \textbf{YW} & \textbf{ITL} & \textbf{TDD} \\ \hline
\textbf{G1} & SS & BSK & MR \\
\textbf{G2} & MR & SS & BSK  \\ 
\textbf{G3} & BSK & MR & SS \\  \hline 
\end{tabular} 
\end{table}

As Table \ref{schedule} shows, Paf's experiment was designed as a two-factor three-level within-subjects experiment \cite{juristo2013basics}. Subjects were assigned to groups (i.e., G1, G2, G3) by means of stratified randomization attending to skills \cite{wohlin2012experimentation}. All groups applied YW, ITL and TDD on the first, second, and third day, respectively. Development approaches were applied in this order so as to minimize learning effects from the least known development approach (i.e., TDD) to the most known development approach (i.e., YW). Each group was assigned to a different combination of tasks so as to balance out the influence of task on results.

Table \ref{tab:settings_baseline_replications} summarizes the settings of the experiment conducted at Paf.

\begin{table}[h!]
\small
\begin{center}
\caption{Paf experiments' settings.}
\label{tab:settings_baseline_replications}
\begin{tabular}{ll} \hline \hline
\textbf{Aspect} & \textbf{Values}  \\ \hline
\textbf{Development Approach} & YW. vs. ITL vs. TDD  \\ 
\textbf{Tasks} & BSK vs. MR vs. SS \\
\textbf{Response variable} & QLTY \\ 
\textbf{Design} & Within-subjects design \\ 
\textbf{Training} & TDD seminar \\ 
\textbf{Training duration} & 3 days/6 hours\\ 
\textbf{Experiment duration} & 2.25 hours\\ 
\textbf{Technological Environment} & Java, Eclipse, JUnit\\ \hline
\end{tabular}
\end{center}
\end{table}

\subsection{Analysis Approach}

First, we provide the descriptive statistics (i.e., mean, standard deviation and median) of YW, ITL and TDD. Then, we complement the descriptive statistics with violin-plots and box-plots to ease the understanding of the data. 

Afterwards, we analyze the experiment with a Linear Mixed Model (LMM) \cite{brown2014applied} following the top-down strategy proposed by West et al. \cite{west2014linear}. In particular, the interaction between Treatment and Task was dropped out as the interaction was not statistically significant. Then, we use pairwise contrasts to convey the difference in performance between development approaches. We used Tukey's correction for multiple comparisons to provide the contrasts \cite{field2013discovering}. LMM's assume that the residuals are normally distributed. The normality assumption of the residuals was checked via the customary Shapiro-Wilk test \cite{field2013discovering}. 

\section{Analysis}
\label{analysis}
\subsection{Descriptive Statistics}

\begin{figure}[h!]
\includegraphics[width=9cm,keepaspectratio]{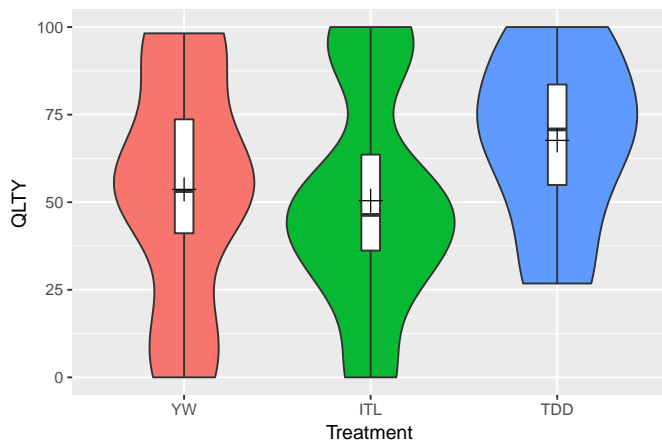}
\caption{Box-plot and violin-plot: YW vs. ITL vs. TDD.} \label{fig1}
\end{figure}

Table \ref{descriptive_statistics} shows the descriptive statistics (i.e., mean, sd, median) of each treatment group (i.e., YW, ITL and TDD). Figure \ref{fig1} shows their corresponding box-plots and violin-plots.

\begin{table}[h!] \centering 
  \caption{Descriptive statistics: YW vs. ITL vs TDD.} 
  \label{descriptive_statistics} 
\begin{tabular}{lcccc} \hline \hline 
\textbf{Treatment} & \textbf{Mean} & \textbf{SD} & \textbf{Median} \\ \hline 
YW & $53.65$ & $34.12$ & $53.18$ \\
ITL & $50.43$ & $32.77$ & $46.37$ \\ 
TDD & $67.64$ & $26.24$ & $70.78$ \\ 
\hline 
\end{tabular} 
\end{table} 

As it can be seen in Table \ref{descriptive_statistics} and Figure \ref{fig1}, TDD scores seem larger and less variable than those of YW and ITL. On the contrary, YW and ITL scores seem to resemble to each other. Not large deviations from normality are expected in the data in view of the data distributions (as all distributions look bell-shaped according to Figure \ref{fig1}).

\subsection{Main Analysis}
A LMM with the main effects of treatment and task was used to analyze the data.\footnote{The lme4 package \cite{bates2014fitting} of the R programming language was used to fit the LMM.} Table \ref{results_lmm} shows the results of the LMM fitted. The Shapiro-Wilk test of the residuals is compatible with the normality assumption ($p$-value=0.119). As it can be seen in Table \ref{results_lmm}, development approaches seem to perform similarly (as the estimates for TDD and ITL are relatively small). The effect of task seems small either, even though the impact of SS ($M=-23.60; SEM\footnote{SEM stands for standard error of the mean. We use SEM and not SE (i.e., standard error) so as not to confuse Software Engineering (defined as SE at the beginning of the article) and standard error.}=18.48$) seems more relevant than that of MR ($M=5.55; SEM=20.24$) on results. 

\begin{table}[h!] \centering 
  \caption{LMM main analysis.} 
  \label{results_lmm} 
\begin{tabular}{lccc} \hline \hline
\textbf{Factor} & \textbf{Estimate} & \textbf{SEM} & \textbf{$p$-value} \\ \hline
Intercept & 65.46 & 14.31 & $<$0.001  \\ 
ITL & -6.00 & 18.48  & 0.749   \\ 
TDD & 0.330 & 18.47 & 0.985   \\ 
MR & 5.548 & 20.24 & 0.787  \\ 
SS & -23.60 & 18.48 & 0.218   \\ \hline
\multicolumn{4}{l}{*Reference class: YW-BSK}
\end{tabular} 
\end{table}

Table \ref{contrasts} shows the pairwise contrasts between development approaches.

\begin{table}[h!] \centering 
  \caption{Pairwise contrasts on treatments.} 
  \label{contrasts} 
\begin{tabular}{lccc} \hline \hline
\textbf{Contrast} & \textbf{Estimate} & \textbf{SEM} & \textbf{$p$-value} \\ \hline 
YW vs. ITL & 6.01 & 18.48 & 0.94  \\ 
YW vs. TDD & -0.33 & 18.48 & 0.99   \\ 
ITL vs. TDD & -6.34 & 20.24 & 0.95   \\ \hline
\end{tabular} 
\end{table}

As it can be seen in Table \ref{contrasts}, YW outperforms ITL to a small extent ($M=6.01; SEM=18.48$). Besides, TDD slightly outperforms YW ($M=-0.33; SEM=18.48$). In view of these findings, YW (i.e., Paf's current development approach) seems to perform slightly worse than TDD and just a bit better than ITL. However, differences are not statistically significant, and thus, could have been observed just by chance.

\section{Comparing Paf's Results and Previous Experiments' Results}
\label{comparing}

\begin{figure}[h!]
\includegraphics[width=9cm,keepaspectratio]{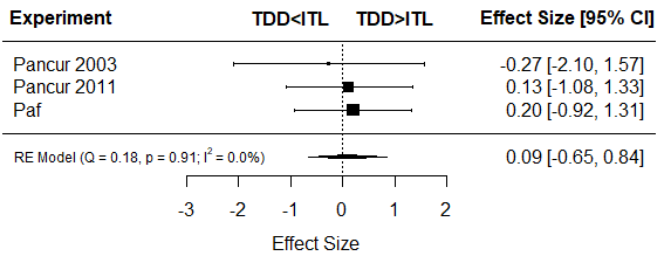}
\caption{Forest-plot: TDD vs ITL.} \label{forest}
\end{figure}

Paf's results cannot be compared with those of any other industrial experiment---as the only experiment conducted so far in industry according to the secondary studies on TDD (i.e., Geras et al. \cite{geras2004prototype}), could not finally assess the extent to which ITL and TDD performed with regard to quality. However, Paf's results can be compared with those of academic experiments. In particular, our results on the performance of ITL and TDD agree with those reported by Pan{\v{c}}ur et al. in 2011 \cite{panvcur2011impact}: TDD outperforms ITL on quality to an almost negligible extent. However, just opposite results (i.e., ITL slightly outperforms TDD on quality) were found by Pan{\v{c}}ur et al. in an earlier academic experiment \cite{pancur2003towards}. To draw a joint conclusion from these results, we combined them by means of a random-effects meta-analysis \cite{borenstein2011introduction}. We used a random-effects meta-analysis as it provides identical results than those provided by a fixed-effects meta-analysis if there was no heterogeneity of results, and thus, should be used by default to aggregate the results from different studies gathered from literature \cite{borenstein2011introduction, whitehead2002meta}. Figure \ref{forest} shows the forest-plot corresponding to the meta-analysis that we performed.

As it can be seen in Figure \ref{forest}, the joint effect size is small (i.e., Cohen's d=$0.09$) and non-statistically significant (i.e., the 95\% crosses 0). In view of the conflicting evidence achieved so far with regard to the performance of ITL and TDD on external quality (2 experiments show that TDD outperforms ITL and viceversa in another), and in view of the small effect sizes, and the small joint effect size, we suggest, \textit{TDD slightly outperforms ITL for TDD novices}. 

\section{Discussion}
\label{discussion}

With regard to RQ1 (i.e., Does TDD outperform current development practices at Paf with regard to external software quality?), \textit{TDD slightly outperformed Paf's current development practices} (i.e., YW) on external software quality. On average, ITL performed the worst during the experiment. Despite the encouraging results obtained with TDD, we could not recommend Paf's developers to \textit{immediately} adapt TDD: not much difference in performance between YW and TDD was observed along the experiment. 

However, as TDD was never applied before by any developer at Paf, and as developers achieved similar results with TDD that with YW, TDD seems a promising development approach at Paf. In view of this, we suggest Paf's managers to consider moving \textit{some} of their developers to TDD, and eventually, after they are acquainted with enough experience, \textit{run again another experiment}. We suggest that ideally, this new experiment should have \textit{two independent groups}: one with experts on TDD and another with experts on YW. This new experiment will not have the shortcoming of evaluating a completely new development approach (i.e., TDD) against a more usual approach (i.e., YW) and thus, risk obtaining favorable results to more  "traditional" approaches. In addition, this new experiment may serve better for making a "go/no-go" decision on whether to adopt TDD at Paf. We do not recommend evaluating the performance of ITL in this new experiment as it was not as effective as TDD, and as studying its performance further may imply dividing developers into more groups, and eventually, to lower the representativeness of the results achieved in the new experiment (as less developers can be assigned to either YW or TDD).

Regarding RQ2 (i.e., Do Paf's results agree with those of previous experiments on TDD?), \textit{Paf's results could not be compared with those of any industrial experiment}---as none allows to evaluate the performance of TDD on external quality. However, \textit{Paf's results seem to agree with those of previous academic experiments}. In particular, a negligible difference in performance between TDD and ITL have also been observed in those. However, TDD outperformed ITL at Paf and at another experiment \cite{panvcur2011impact}, but not at another one \cite{pancur2003towards}. Natural variation of results may be behind the differences of results observed across the experiments. Particularly, as experiments have small sample sizes (15 subjects in our experiment, 32 in \cite{panvcur2011impact} and 38 in \cite{pancur2003towards}), if in reality the difference in performance between ITL and TDD was small (as it seems the case), just by chance, negative and positive results may emerge \cite{button2013power}. This may explain why experiments show conflicting results with regard to the performance of TDD on quality. 

Finally, after combining the results of all the experiments together by means of meta-analysis, TDD slightly outperformed ITL. In view of the almost negligible benefit obtained with TDD in controlled experiments, and the large benefits obtained with TDD in case studies and surveys, we hypothesize, \textit{the lack of previous familiarity of the experiments' participants with the TDD process may have impacted results}. In view of this, we hypothesize that TDD may not show its full potential until developers are already acquainted with enough practice, and that this may be the reason behind the drop of quality observed in most experiments with regard to more "traditional" approaches. However, this finding should be further substantiated, as yet there is a scarce number of industrial experiments evaluating the effects of TDD on external software quality.

\section{Threats to Validity}
\label{threats}
In this section, we report the main threats to the validity of our study following Wohlin et al.'s recommendations \cite{wohlin2012experimentation}. The validity threats are prioritized according to Cook and Campbell's guidelines \cite{cook1979quasi}.

\textbf{Conclusion validity} concerns the statistical analysis of results \cite{wohlin2012experimentation}. We provide numerical evidence on the validity of the required statistical assumptions of the statistical model used (i.e., Linear Mixed Model \cite{brown2014applied}). Particularly, we assess the normality of the data with the Shapiro-Wilk test, a commonly used statistical test in SE \cite{wohlin2012experimentation}. The random heterogeneity of the sample threat might have materialized in our experiment since professionals had different levels of experience. This might have biased the results towards the average performance (i.e., the performance of seniors and juniors), thus resulting in non-significant results.

\textbf{Internal validity} is the extent to which the observed effects are caused by the treatments and not by other variables beyond researchers' control \cite{wohlin2012experimentation}. There is a potential maturation threat: the course was a three-day seminar on TDD and contained multiple exercises and laboratories. Thus, factors such as tiredness or inattention might have materialized. In order to minimize this threat, we offered professionals the choice of the schedule that best suited their needs, and we ensured that subjects were given enough breaks. Training leakage may have distorted results. Even though training leakage could not materialize in the first two sessions (as in the first session no training was required and in the second just training on ITL was given), this was a possibility in the third session (as subjects were already knowledgeable on ITL and TDD). Particularly, subjects may have learned something during the ITL session (e.g., how to develop in micro-iterative steps) that may have boosted their performance during the TDD session. However, we do not think this threat materialized as the difference in performance between sessions was almost negligible. There was also the possibility of a diffusion threat: since subjects performed different development tasks in each experimental session, they could compare notes at the end of the sessions and thus, aid their colleagues to boost their performance in sub-sequent sessions. This could lead to an improvement in their performance. To mitigate this threat, we encouraged subjects not to share any information with their colleagues until the end of the three-day training course. 

\textbf{Construct validity} refers to the correctness in the mapping between the theoretical constructs and the operationalizations of the variables in the study \cite{wohlin2012experimentation}. As usual in SE experiments, Paf's experiment suffers from mono-operation bias (external quality was just measured with test cases). Conformance to the development approaches is one of the big threats to construct validity of SE experiments. However, this threat to validity was minimized by visual supervision and by encouraging subjects to adhere as closely as possible to the development approaches taught during the seminar. There were no significant social threats, such as evaluation apprehension: all subjects participated on a voluntary basis in the experiment and they were ensured that their data were going to be treated anonymously.

\textbf{External validity} relates to the possibility of generalizing results beyond the objects and subjects of the study \cite{wohlin2012experimentation}. As usual in SE experiments, our experiment was exposed to the selection threat since we did not have the opportunity to randomly select subjects from a population; instead, we had to rely on convenience sampling. Java was used as the programming language during the experimental sessions and the measurement of the outcomes. This way, we addressed possible threats regarding the use of different programming languages to measure the outcomes. However, this limits the validity of our results to this language only. The three tasks used in the experiment (i.e., BSK, MR and SS) were toy tasks. This affects the generalizability of the results and their applicability in industrial settings. The task domain might not be representative of real-life applications, and the duration of the experiment (two hours and 15 minutes to perform each task) might have had an impact on results. We acknowledge that this might be also an obstacle to the generalizability of the results. However, we expect our results to be representative for professionals starting to learn the TDD process with toy-tasks.

\section{Conclusions}
\label{conclusion}

TDD has been claimed to increase external software quality compared to traditional development approaches (e.g., Waterfall or ITL) across numerous studies \cite{kollanus2010test, shull2010we, causevic2011factors, rafique2013effects, makinen2014effects, munir2014considering, bissi2016effects}. Unfortunately, almost none of them is an industrial experiment. In view of the encouraging results achieved with TDD, Paf's managers were interested in evaluating the extent to which TDD would perform at their premises.

We ran an experiment at Paf to evaluate the extent to which YW (i.e., the traditional way of coding at Paf), ITL and TDD would perform on external software quality. All development approaches performed similarly (i.e., differences in results could have been observed just by chance). Thus, we could not recommend Paf's managers to immediately move their developers to TDD. However, in view that TDD seems a promising approach at Paf (as without having applied TDD before, developers achieved similar results to those obtained with YW), we recommend Paf's managers to move some of their developers to TDD and eventually, after developers are acquainted with enough experience, to run a new experiment comparing the performance of TDD and YW. This new experiment may serve to make a "go/no-go" decision on whether to adopt TDD at Paf. 

The results obtained at Paf cannot be compared with those of any other industrial experiment (as the only experiment conducted so far on TDD could not evaluate its effects on quality \cite{geras2004prototype}). However, Paf's results can be compared with those achieved in academic experiments. In view of our results and those, ITL and TDD behave similarly in terms of external software quality for novice developers on TDD coding toy-tasks. In addition, experiments' small sample sizes may be behind the observed differences across experiments' results. Specifically, differences across experiments' results (i.e., TDD effects are positive in two experiments and negative in another one) may emerge due to the variation of results expected in small sample sizes, and the plausible presence of a real small difference in performance \cite{button2013power}.

Finally, while case studies and surveys show large improvements with TDD on external quality, this seems not supported by experiments. Differences between experiments' results and those of case studies and surveys may be due to the lack of previous familiarity of experiments' participants with the TDD process. In particular, TDD may not show its full potential until enough experience has been gained with it, and this may be the reason behind the differences of results observed across research methods. However, this finding should be further substantiated, as yet there is a scarce number of industrial experiments evaluating the performance of TDD on external software quality.

As a concluding remark, despite the long years of research on TDD, we have not been able yet to obtain definite results on TDD's performance in experimental settings---particularly in industry, where the lack of experiments seems more pronounced. We---as well as other researchers did before \cite{offutt2018don}---encourage SE researchers to continue investigating the effects of TDD in industrial relevant scenarios, so eventually, the performance of TDD in daily practice can be understood.

\section*{Acknowledgments}
We would like to thank to all the participants in the experiment and to Paf's managers: this research would have not been possible without your help. This research was developed with the support of the Spanish Ministry of Science and Innovation project TIN2014-60490-P.

\bibliographystyle{IEEEtran}
\bibliography{IEEEabrv,biblio}

\end{document}